\documentclass[sigconf]{acmart}

\copyrightyear{2025}
\acmYear{2025}
\setcopyright{cc}
\setcctype{CC-BY}
\acmConference[SAC '25]{The 40th ACM/SIGAPP Symposium on Applied
Computing}{March 31-April 4, 2025}{Catania, Italy}
\acmBooktitle{The 40th ACM/SIGAPP Symposium on Applied Computing (SAC'25), March 31-April 4, 2025, Catania,
Italy}\acmDOI{10.1145/3672608.3707817}
\acmISBN{979-8-4007-0629-5/25/03}

\begin{document}
\title{The Use of Generative Artificial Intelligence for Upper Secondary Mathematics Education Through the Lens of Technology Acceptance}
  
\renewcommand{\shorttitle}{GenAI in Upper Secondary Math Education}

\author{Mika Setälä}
\orcid{0009-0006-7195-9587}
\affiliation{%
  \institution{University of Jyväskylä}
  \department{Faculty of Information Technology}
  \city{Jyväskylä} 
  \country{Finland}
}

\author{Ville Heilala}
\orcid{0000-0003-2068-2777}
\affiliation{%
  \institution{University of Jyväskylä}
  \department{Faculty of Education and Psychology}
  \city{Jyväskylä} 
  \country{Finland}
}

\author{Pieta Sikström}
\orcid{0000-0002-2055-7995}
\affiliation{%
  \institution{University of Jyväskylä}
  \department{Faculty of Information Technology}
  \city{Jyväskylä} 
  \country{Finland}
}

\author{Tommi Kärkkäinen}
\orcid{0000-0003-0327-1167}
\affiliation{%
  \institution{University of Jyväskylä}
  \department{Faculty of Information Technology}
  \city{Jyväskylä} 
  \country{Finland}
}

\renewcommand{\shortauthors}{Setälä et al.}

\begin{abstract}
This study investigated the students' perceptions of using Generative Artificial Intelligence (GenAI) in upper-secondary mathematics education. Data was collected from Finnish high school students to represent how key constructs of the Technology Acceptance Model\textemdash Perceived Usefulness, Perceived Ease of Use, Perceived Enjoyment, and Intention to Use\textemdash influence the adoption of AI tools. First, a structural equation model for a comparative study with a prior study was constructed and analyzed. Then, an extended model with the additional construct of Compatibility, which represents the alignment of AI tools with students' educational experiences and needs, was proposed and analyzed. The results demonstrated a strong influence of perceived usefulness on the intention to use GenAI, emphasizing the statistically significant role of perceived enjoyment in determining perceived usefulness and ease of use. The inclusion of compatibility improved the model's explanatory power, particularly in predicting perceived usefulness. This study contributes to a deeper understanding of how AI tools can be integrated into mathematics education and highlights key differences between the Finnish educational context and previous studies based on structural equation modeling.
\end{abstract}

%
%

\ccsdesc[500]{Computing methodologies~Artificial intelligence}
\ccsdesc[500]{Applied computing~Computer-assisted instruction}
\ccsdesc[300]{Human-centered computing~Human computer interaction (HCI)}

\keywords{mathematics education, upper secondary school, generative artificial intelligence, technology acceptance}

\maketitle

\section{Introduction}

Education is constantly evolving, and technological innovations continually provide new opportunities to enhance teaching and learning practices \cite{valtonen2022nature}. In the 1990s, graphic calculators or CAS (Computer Algebra System) were introduced in education and expected to revolutionize the teaching and learning of mathematics in upper secondary education. However, at the same time, concerns arose about how students would be motivated to train traditional calculation skills when an efficient calculator could perform calculations for them easily and quickly   \cite{PenglaseArnold1996}. In fact, graphical calculators enhanced students' sense of control by allowing them to simultaneously check the accuracy of their solutions \cite{Joutsenlahti2005}. When students used graphic calculators in both learning and exams, their ability to perform calculations, understand the basic ideas of mathematics, and achieve better learning results improved \cite{Ellington2006}.

In general, using graphic calculators in mathematics education led to a shift towards more inquiry-based learning, as these calculators could quickly illustrate mathematical concepts \cite{Joutsenlahti2005}. Despite the concerns, graphic calculators were approved for use in Finland's matriculation examination (the national school-leaving exam for general upper secondary education) in the early 1990s. This decision allowed students to use a broader range of mathematical formulas and shifted the focus from memorizing formulas to meaningfully applying them.

To date, graphic calculators have become standard tools in mathematics education, but the increased use of Generative Artificial Intelligence (GenAI) techniques utilizing large language models is currently sparking similar concerns and discussions as calculators did in the 1990s. Teachers and students are interested in GenAI's potential to support student learning and reduce teachers' workload, but at the same time, they are concerned about the effects on learning outcomes \cite{sikstrom2024pedagogical}. 
Due to conversational abilities, GenAI can act as a pedagogical agent scaffolding learning and supporting students' intrapersonal learning processes, such as self-regulation, motivation, and meta-cognition \cite{sikstrom2022pedagogical}. For example, in mathematics education, GenAI utilizing large language models fundamentally enlarges possibilities to get instant individual feedback \cite{liu2023summary,naveed2023comprehensive,Heilala2024-be}. This kind of personalized learning can be supported by GenAI responding to math queries and serving as a search engine and knowledge base for students \cite{frieder2023mathematical}. 

The role and use of AI tools in education largely depend on their readiness, impact, and benefits, all of which need to be carefully evaluated before integrating them into teaching and learning practices \cite{bond2024meta}. Therefore, to ease the adoption of GenAI in learning, it is essential to investigate students' perceptions of the usefulness of new AI tools. This is also important because both teachers and students need to develop new skills to effectively understand and utilize these technologies \cite{Heilala2023-jh,kasneci2023chatgpt,sikstrom2024pedagogical}. 

Due to the recent emergence of GenAI and the continuous development of large language models \citep[e.g.,][]{Heilala2024-be}, the educational field needs empirical results from real-world experiences. Thus, the present study, conducted in Finland, addresses the research gap by exploring how students perceive using GenAI in advanced mathematics in general upper secondary education. 

The study intends to compare and extend prior research by drawing inspiration from the work by \citet{lai2023}, which utilized the Technology Acceptance Model (TAM) and partial least squares structural equation modeling (PLS-SEM) to examine ChatGPT adoption in higher education in Hong Kong. Specifically, this study's aim is to reproduce and extend prior work, which involves a different researcher analyzing their own dataset using an analysis method tailored to the study, to examine the hypotheses established in the baseline research \citep{Ihantola2015-pq}.
 
This study employs the constructs from the TAM model to explore students' perceptions of using AI-based educational tools. First, the key constructs measured in this study include Perceived Usefulness (PU), Perceived Ease of Use (PEOU), Perceived Enjoyment (PE), and Intention to Use (ITU). In addition, the study compares the results of both the Finnish and Hong Kong studies \cite{lai2023} to identify differences in students' insights and perceptions. Second, in contrast to \cite{lai2023} focusing solely on these established TAM variables, this research extends the model by introducing Compatibility (COM), a variable reflecting the degree to which the use of the AI tool aligns with students' prior experience, beliefs, and educational needs. The inclusion of Compatibility provides a nuanced perspective on how well these technologies align with students' pre-existing digital skills and educational expectations. 

The results of the study are expected to offer valuable and practical insights into how students perceive the integration of GenAI, thereby supporting teachers in their decision-making about incorporating GenAI into their teaching. Thus, the central research questions are as follows: 1) To what extent do key constructs of the TAM model influence students' Intention to Use GenAI in general upper secondary mathematics education? 2) Are there differences in how these constructs operate in the Finnish educational context compared to findings from similar studies? 3) How does the additional variable, Compatibility, affect students' perceptions of using GenAI tools?

By answering these research questions, the study aims to broaden and deepen the understanding of how both traditional TAM variables and the additional factor, Compatibility, reflect the adoption and integration of AI tools, such as Copilot, in general upper secondary mathematics education. 

\section{Background on technology acceptance in mathematics education}

The theoretical framework used in this study is the Technology Acceptance Model (TAM), which aims to explain how users come to accept and use technology, focusing on two primary factors: perceived usefulness (PU) and perceived ease of use (PEOU), which influence users' intention to use the technology (ITU) \cite{Davis1989-rx,Davis2024-gz}. During later stages, the TAM model has been refined to incorporate additional variables, and other researchers have applied and suggested various extensions to the TAM \citep{granic2019technology}. Thereby, the original TAM model has been extended with factors such as social influence and facilitating conditions in the follow-up models TAM2 and the unified theory of acceptance and use of technology (UTAUT) \cite{venkatesh2000determinants, venkatesh2000theoretical}.
TAM and its extensions have also been used in mathematics education to examine university students' perceptions of technology use, students' behavioral intentions and achievements in ICT-integrated mathematics instruction, and the use of augmented reality in project-based geometry learning \cite{zogheib2015,chen2020students,johar2021examining}. For example, in upper-secondary education in Norway, the main motivation of the students to use graphic calculator equipment was its efficiency, enabling them to solve procedural problems more quickly \cite{mohammad2019}. These findings align with broader studies, \citet{zogheib2015} and \citet{song2017}, indicating that students' acceptance of technological tools in mathematics is strongly influenced by perceived usefulness and ease of use. Perceived enjoyment is suggested to influence key elements of the TAM, especially perceived usefulness and ease of use, highlighting that users are more likely to find the technology useful and easy to use when they enjoy it \cite{Teo2011-zp}.

The study by \citet{lai2023} explored the adoption of ChatGPT as a tool to support active learning in higher education through the extended TAM, where the construct of intrinsic motivation (IM) was added to TAM. The research, conducted among 473 undergraduate students in Hong Kong, sought to investigate how intrinsic motivation influences the intention to use ChatGPT in educational settings. Using structural equation modeling, the study confirmed that intrinsic motivation is the strongest predictor of students' behavioral intention to adopt ChatGPT. While perceived usefulness remains a significant factor in predicting technology adoption, consistent with existing TAM literature, the study diverged from previous findings by showing no significant impact of perceived ease of use on students' intention to use ChatGPT. Furthermore, the results revealed that neither Perceived Usefulness nor Ease of Use acted as mediators between intrinsic motivation and behavioral intention. This suggests that while the practical benefits of ChatGPT are acknowledged, its adoption in education is more heavily influenced by students' inherent desire for engaging and stimulating learning experiences. 

\citet{strzelecki2023} investigated the acceptance and usage of ChatGPT in higher education using the UTAUT2 model. The study, conducted with 534 students from a Polish university, sought to determine the factors influencing behavioral intention and use of ChatGPT. Seven key predictors were used, including Performance Expectancy (corresponds to Perceived Usefulness and Effort Expectancy), which aligns with Perceived Ease of Use. Additionally, Hedonic Motivation (similar to Perceived Enjoyment), was evaluated to understand the role of enjoyment in technology adoption. The results revealed that Habit was the most significant predictor of Behavioral Intention, with Performance Expectancy and Hedonic Motivation also playing crucial roles. In contrast, Effort Expectancy and Social Influence had smaller effects on students' intention to use ChatGPT, indicating that while ease of use is important, it is not as influential as the perceived usefulness or enjoyment of the tool. The study highlights that intrinsic motivators such as habit and enjoyment strongly predict technology adoption.

\begin{table*}[!htbp]
\centering
\caption{Details of the teaching experiments from 2023 to 2024.}
\label{table:combined_experiments}
\begin{tabular}{p{0.18\textwidth} p{0.18\textwidth} p{0.18\textwidth} p{0.18\textwidth} p{0.18\textwidth}}
\toprule
\textbf{Category} & \textbf{Third-year students (Autumn 2023)} & \textbf{First-year students (Autumn 2023)} & \textbf{Second-year students (Spring 2024, Logic and Programming)} & \textbf{Second-year students (Spring 2024, Congruence)} \\ 
\midrule
Participants & $n=19$& $n=45$& $n=33$& $n=38$\\ 
\hline
Topic & Normal distribution and normalization & Written assignments on equations and systems of equations & Logic and programming & Congruence and Diophantine equations \\ 
\hline
Duration & 2 x 75 min lessons, 2 homework sets & 75 min lesson & 3 x 75 min lessons, 3 homework sets & 3 x 75 min lessons, 3 homework sets \\ 
\hline
Tasks & Electronic textbook exercise & Electronic textbook exercises & Electronic textbook exercise & Electronic textbook exercises \\ 
\hline
Use of AI & Teacher-prepared prompts given & Teacher-prepared prompts given & Self-directed prompting & Self-directed prompting \\ 
\bottomrule
\end{tabular}
\end{table*}

\section{Materials and methods}

\subsection{Participants, research design and pedagogical approach}

The research was conducted in a Finnish general upper secondary school (ISCED level 3). As depicted in Table \ref{table:combined_experiments}, four teaching experiments with different characteristics were conducted during the 2023--2024 semester in which a total of 135 participants used GenAI. In the experiments, students studied with the assistance of \textsc{Copilot}, which was tested for stable operation and was available for the students using their school-provided user accounts. 

During autumn 2023, two study sessions were held for the advanced mathematics (i.e., advanced mathematics is a more rigorous and extensive track compared to basic mathematics) courses: one for the first-grade students (16--17 years old) and the other for the third-grade students (18–-19 years old). Among the first-grade students, 45 used an AI agent to assist with verbal problem-solving exercises involving equations, starting directly with the exercises without any prior instruction. In contrast, 19 third-grade students who used the AI agent began their session with a collective lesson on normal distribution. After the lesson, they completed exercises using either teacher-prepared prompts or their own prompts to interact with the AI agent, similar to the first-grade group. 

Finally, during spring 2024, two study sessions were conducted for the Algorithms and Number Theory course in advanced mathematics for second-grade students (17-–18 years old). In the first session, 33 students studied the topic of logic and programming using GenAI. In the second session, 38 students explored congruence and Diophantine equations, again utilizing GenAI to assist with the exercises.
 
\subsection{Instruments and data analysis}

This study utilizes the TAM model with the core constructs of perceived usefulness (PU), perceived ease of use (PEOU), and intention to use (ITU)  \cite{Davis1989-rx,Davis1989-dw,Davis2024-gz}. PU refers to an individual's belief that using a specific system will improve their job performance, suggesting a positive link between a system's use and the work outcomes \cite{Davis1989-dw}. PEOU refers to the belief that using a system requires minimal effort, and systems perceived as easier to use are more likely to be accepted by users \cite{Davis1989-dw}.  ITU depicts the likelihood that an individual will engage in using a specific technology in the future, which is often considered a key predictor of actual technology adoption and usage \citep{Davis1989-rx, Davis2024-gz}.

The TAM model can be extended and modified by incorporating additional constructs, often simply by adding one or two elements \citep{benbasat2007quo}. In this study, the TAM model has been extended with two additional constructs to better understand how \textit{enjoyment} and \textit{compatibility} influence the core constructs of the model. Usefulness and enjoyment have been shown to mediate fully the effects on usage intentions of perceived usefulness and perceived ease of use \cite{Davis1992-af}. Thus, perceived enjoyment (PE), the intrinsic pleasure derived from using technology, has been added in this study as  an important factor influencing user acceptance and intention to use technology.  Second, the extension of TAM in this study includes Compatibility (COM), which refers to the degree of harmony between the user's general beliefs and a system \cite{rogers2014diffusion}. All these constructs are presented in Table \ref{table:affectiveinstruments}.

\begin{table*}[!htbp]
\centering
\caption{Affective measurement instruments used in the study}
\begin{tabular}{p{0.3\textwidth} p{0.6\textwidth}}
\toprule
\textbf{Measurement Instrument} & \textbf{Definition} \\
\hline
Perceived Usefulness (PU) & The degree to which a person believes that using a particular system would enhance their job (learning)  performance \cite{Davis1989-dw,Davis2024-gz}\\
Perceived Ease of Use (PEOU) & The degree to which a person believes that using the system would be free of effort \cite{Davis1989-dw,Davis2024-gz} \\
Perceived Enjoyment (PE) & The extent to which the activity of using the system is perceived to be enjoyable in its own right, apart from any performance consequences that may be anticipated \cite{Davis1992-af,Davis2024-gz} \\
Compatibility (COM)& The degree to which an innovation is in harmony with the user's beliefs, knowledge, experience, and existing requirements \cite{rogers2014diffusion}\\
Intention to Use (ITU)& The likelihood that an individual will engage in using a specific technology in the future, which is considered a direct predictor of actual technology usage \cite{Davis1989-rx,Davis2024-gz}\\
\end{tabular}
\label{table:affectiveinstruments}
\end{table*}

The internal consistency of these localized constructs was evaluated using coefficient $\alpha$. All constructs exhibited high reliability, with $\alpha > 0.85$, indicating excellent internal consistency. Structural Equation Modeling (SEM) is a statistical technique that can be used to analyze and understand complex relationships between variables. It integrates aspects of factor analysis and multiple regression to study both observed and latent variables and their causal relations simultaneously. The Partial Least Squares SEM (PLS-SEM) has been widely used in education research \cite{Ghasemy2020-xv}, for example, to explore the adoption of metaverse technology in engineering education \cite{Wiangkham2024-my} and to predict the actual use of mobile learning systems \cite{Alshurideh2020-ho}.

In this study, the PLS-SEM analysis was conducted using the \textit{seminr} package in R. The \textit{seminr} package provides a flexible framework for defining both measurement and structural models, allowing for the estimation of complex relationships between latent variables. The PLS-SEM model was estimated by first specifying the measurement model for constructs such as PU and ITU, followed by the structural model, which defines the relationships between these constructs. The structural model was assessed with 5,000 bootstrap resamples to evaluate the significance of path coefficients. 

\begin{table*}[t!]
\centering
\caption{Reliability Analysis of Constructs}
\begin{tabular}{lccc}
\toprule
\textbf{Construct} & \textbf{Alpha} & \textbf{Standardized Alpha} & \textbf{Item-Total Correlation (Range)} \\
\hline
Perceived Usefulness (PU) & 0.91 & 0.91 & 0.76 -- 0.85 \\
Perceived Ease of Use (PEOU) & 0.85 & 0.86 & 0.62 -- 0.75 \\
Perceived Enjoyment (PE) & 0.92 & 0.92 & 0.79 -- 0.84 \\
Compatibility (COM) & 0.87 & 0.87 & 0.66 -- 0.81 \\
Intention to Use (ITU) & 0.93 & 0.93 & 0.82 -- 0.85 \\
\bottomrule
\end{tabular}
\end{table*}

Before conducting the main analysis, several preliminary checks were performed to ensure the dataset's suitability for PLS-SEM. The sample size was determined using the 10-times rule and a power analysis. The 10-times rule required a minimum of 40 respondents based on the most complex path involving ITU with three predictors (PU, PEOU, PE) and four indicators. The power analysis, assuming a medium effect size ($f^2 = 0.15$), a significance level ($\alpha = 0.05$), and statistical power ($1 - \beta = 0.80$), indicated that the sample size was sufficient \cite{hair2016primer}. Thereby, the final sample size of 102 respondents was sufficient to provide reliable and statistically significant results. No missing data were found across the variables.
Descriptive statistics indicated that skewness (ranging from -0.26 to 0.22) and kurtosis (ranging from -1.04 to -0.68) for all observed variables were within acceptable ranges, suggesting no severe departures from normality \cite[e.g.,][]{hair2016primer}. Thirteen outliers were identified using the Mahalanobis distance, but these were considered legitimate. Multicollinearity was assessed using the Variance Inflation Factor (VIF), with most values below five within sufficient limits \cite{Hair2019-gj}. However, four values exceeded 5 (e.g., PE01: 6.1, ITU03: 5.7), indicating moderate multicollinearity, which remains acceptable to a guideline suggesting a threshold of 10 \cite{Cohen2013-tw}. 
The reliability and convergent validity of all constructs were confirmed. The composite reliability (CR) for \textit{PU}, \textit{PEOU}, \textit{PE}, \textit{COM} and \textit{ITU} exceeded the recommended threshold of 0.70 \cite{Hair2019-gj}, with values ranging from 0.903 to 0.948.

\section{Results}

\subsection{Use of existing structural equation model}

\subsubsection{The model}

\begin{figure*}
    \centering
    \includegraphics[width=1\linewidth]{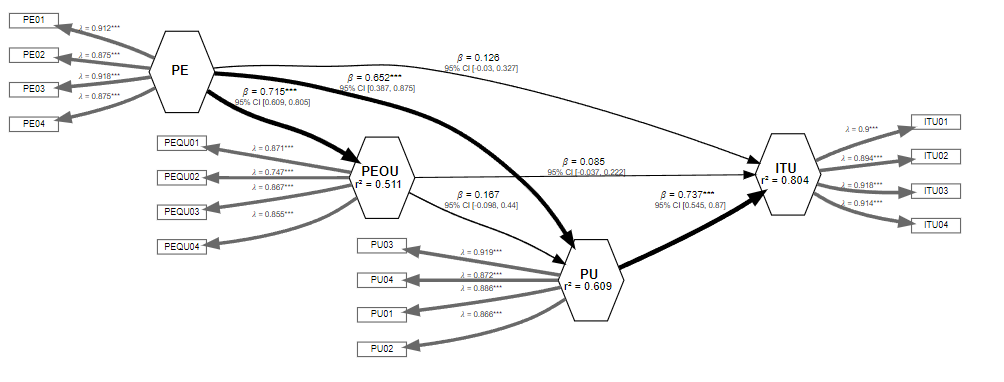}
    \caption{Structural Equation Model (SEM) for the adoption of Generative AI in upper secondary mathematics education based on the Technology Acceptance Model (TAM).}
    \label{fig:sem1}
    \Description[Figure]{Structural Equation Model (SEM) for the adoption of Generative AI in upper secondary mathematics education based on the Technology Acceptance Model (TAM).}
\end{figure*}

The analysis revealed several significant relationships between the constructs PE, PEOU, PU, and ITU (Figure \ref{fig:sem1}) when using the same structural equation model as in the study with students from Hong Kong \cite{lai2023}. 

In particular, PU was strongly influenced by both PE ($\beta = 0.652$, $p < .001$, 95\% CI [0.379, 0.877]) and PEOU ($\beta = 0.167$, $p = .233$, 95\% CI [-0.094, 0.447]), although the latter was not statistically significant. Furthermore, PU had a strong and significant effect on ITU ($\beta = 0.737$, $p < .001$, 95\% CI [0.544, 0.868]).

\textit{PE} also significantly predicted \textit{PEOU} ($\beta = 0.715$, $p < .001$, 95\% CI [0.607, 0.806]) and had a weaker and non-significant direct effect on \textit{ITU} ($\beta = 0.126$, $p = .162$, 95\% CI [-0.030, 0.329]). The path from \textit{PEOU} to \textit{ITU} was also non-significant ($\beta = 0.085$, $p = .201$, 95\% CI [-0.037, 0.227]). 

Overall, the model explained 80.4\% of the variance in \textit{ITU}; 60.9\% of the variance in \textit{PU}, and 51.1\% of the variance in \textit{PEOU}, indicating strong predictive power for the intention to use GenAI in education. 

\subsubsection{Comparison}

In this section, the present study results are compared with the study by \textbackslash{}citet\{lai2023\} to identify differences in the perceptions of Finnish and Hong Kong students. The TAM model was utilized in both studies; however, it is noteworthy that some nominal differences were present referring to the same phenomenon. Thus, for clarification, the \tableautorefname {4} presents the terms corresponding to one another. 

\begin{table*}[t!]
\centering
\caption{Comparison of Constructs Between Finnish and Hong Kong Studies}
\begin{tabular}{p{0.5\textwidth} p{0.5\textwidth}}
\toprule
\textbf{Constructs in Finnish Study} & \textbf{Constructs in Hong Kong Study} \\
\midrule
Perceived Usefulness (PU) & Perceived Usefulness (PU) \\
Perceived Ease of Use (PEOU) & Perceived Ease of Use (PEOU) \\
Perceived Enjoyment (PE) & Intrinsic Motivation (IM) \\
Intention to Use (ITU) & Behavioral Intention (BI) \\
\bottomrule
\end{tabular}
\label{table:constructcomparison}
\end{table*}

In the Hong Kong study, behavioral intention (ITU) was primarily influenced by intrinsic motivation (PE), with a path coefficient of 0.236 (p < 0.001), while the effect of perceived usefulness (PU) was much weaker, with a path coefficient of 0.114 (p = 0.011). In contrast, the Finnish study revealed the opposite pattern: ITU was most strongly driven by PU, with a path coefficient of 0.737 (p < 0.001), whereas the effect of PE was negligible and statistically insignificant (path coefficient = 0.126, p = 0.162). Thus, both studies emphasized one dominant factor influencing ITU, but in Hong Kong, intrinsic motivation was key, while in Finland, the practical utility of the technology played a central role. 

In Hong Kong, PU was influenced by PEOU and PE, but the effects were relatively weak. The path coefficient of PE was 0.091 (p = 0.053), showing a slight but almost significant influence, while the effect of PEOU on PU was minimal (path coefficient = 0.068, p = 0.172). In Finland, PE had a much stronger impact on PU, with a path coefficient of 0.652 (p < 0.001), underlining the importance of intrinsic motivation in shaping perceived usefulness. On the other hand, the effect of PEOU on PU in Finland was also relatively weak and statistically insignificant (path coefficient = 0.167, p = 0.233). 

In both studies, PE had a clear impact on PEOU, but the magnitude of the effect differed significantly. In Hong Kong, the effect of PE on PEOU was relatively small, with a path coefficient of 0.152 (p = 0.001). In Finland, however, the effect was much stronger, with a path coefficient of 0.715 (p < 0.001).

\begin{table*}[!ht]
\centering
\caption{Comparison of Path Coefficients and p-values between HK and FIN studies}
\begin{tabular}{p{0.3\textwidth} p{0.2\textwidth} p{0.1\textwidth} p{0.2\textwidth} p{0.1\textwidth}}
\toprule
\textbf{Hypotheses} & \textbf{HK Path Coefficient} & \textbf{HK p-values} & \textbf{FIN Path Coefficient} & \textbf{FIN p-values} \\
\hline
PE(IM)→ PEOU& 0.152                             & 0.001                      & 0.715                           & 0.000                   \\
PE (IM) → ITU (IB)& 0.236                             & 0.000                      & 0.126& 0.162\\
PEOU → PU           & 0.068                             & 0.172                      & 0.167                           & 0.233                   \\
PEOU → ITU (BI)& 0.042                             & 0.359                      & 0.085                           & 0.201                   \\
PE(IM)      → PU& 0.091                             & 0.053                      & 0.652                           & 0.000                   \\
PU→       ITU (BI)& 0.114                             & 0.011                      & 0.737                           & 0.000                   \\
\bottomrule
\end{tabular}
\label{table:comparison}
\end{table*}


\subsection{Extending the model with compatibility}

\begin{figure*}
    \centering
    \includegraphics[width=1\linewidth]{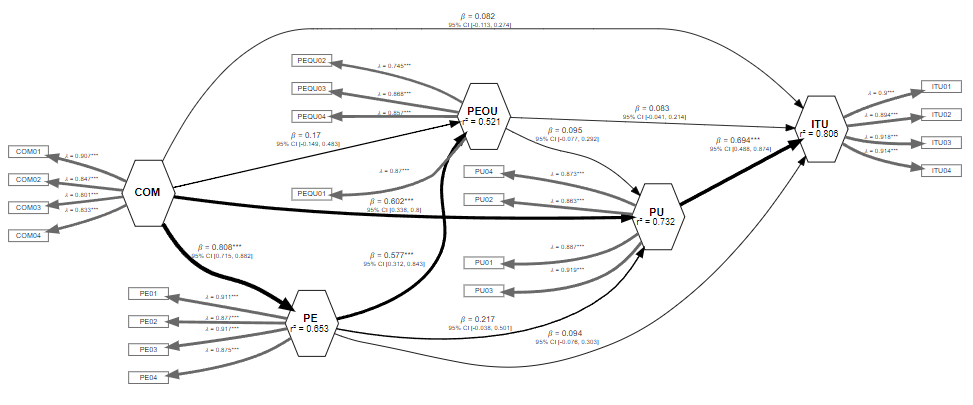}
    \caption{Extended Structural Equation Model (SEM) incorporating Compatibility as an additional factor in the Technology Acceptance Model (TAM).}
    \label{fig:sem2}
    \Description[Figure]{Extended Structural Equation Model (SEM) incorporating Compatibility as an additional factor in the Technology Acceptance Model (TAM).}
\end{figure*}

The extended model (Figure \ref{fig:sem2}), which introduced the COM by \citet{rogers2014diffusion}, was also tested using PLS-SEM and bootstrapping with 5000 resamples.  The relationship between \textit{PE} and \textit{PEOU} (\(\beta = 0.577\), 95\% CI [0.314, 0.843], \textit{p} < .001) was statistically significant, suggesting a strong influence of \textit{PE} on \textit{PEOU}. The path from \textit{COM} to \textit{PU} (\(\beta = 0.602\), 95\% CI [0.335, 0.800], \textit{p} < .001) was also significant, indicating that \textit{COM} is a significant predictor of \textit{PU}. The path from \textit{PU} to \textit{ITU} (\(\beta = 0.694\), 95\% CI [0.478, 0.878], \textit{p} < .001) was highly significant, reinforcing the importance of \textit{PU} as a predictor of \textit{ITU}. However, the paths from \textit{PEOU} to \textit{PU} (\(\beta = 0.095\), \textit{p} = .318) and from \textit{PE} to \textit{ITU} (\(\beta = 0.094\), \textit{p} = .326) were not significant.

\begin{table}[!ht]
\centering
\caption{Summary of Bootstrapped Path Coefficients for the Expanded Model}
\begin{tabular}{lccc}
\hline
Path & Original Estimate & 95\% CI & Significance \\
\hline
PE $\rightarrow$ PEOU & 0.577 & [0.314, 0.843] & *** \\
COM $\rightarrow$ PU & 0.602 & [0.335, 0.800] & *** \\
PU $\rightarrow$ ITU & 0.694 & [0.478, 0.878] & *** \\
PEOU $\rightarrow$ PU & 0.095 & [-0.080, 0.292] & n.s. \\
COM $\rightarrow$ ITU & 0.082 & [-0.113, 0.272] & n.s. \\
\hline
\multicolumn{4}{l}{Note: *** p < .001, n.s. = not significant.} \\
\end{tabular}
\end{table}

Overall, the expanded model with the inclusion of \textit{COM} demonstrates strong predictive power as a significant predictor of both \textit{PU} and \textit{PE}, while certain paths, such as from \textit{PEOU} to \textit{ITU}, were found to be less significant. Adding the COM variable improved the model's explanatory power for PU (R² increased from 0.609 to 0.732) and PEOU (R² from 0.511 to 0.521), with minimal change for ITU (0.804 to 0.806). However, the path from COM to ITU was not significant, indicating limited impact on ITU despite improvements for PU and PE. Including the COM variable introduces a new dimension to the model, significantly impacting both PE (path coefficient = 0.808) and PU (path coefficient = 0.602). This addition broadens the understanding of how COM influences key constructs, such as perceived usefulness and perceived ease of use, making the model more comprehensive. Although the impact of COM on ITU remains minimal, its strong relationship with PE and PU highlights its relevance and importance in explaining these constructs in the context of educational technology and GenAI adoption.

Also, the covariance-based structural equation modeling (CB-SEM) was used to assess the fit of two models (without the COM variable and with the COM variable). The results showed that both covariance-based models fit the data well: without the COM variable, RMSEA was 0.061, CFI 0.999, and SRMR 0.051, while with the COM variable, RMSEA improved slightly to 0.049 while CFI remained at 0.999 and SRMR at 0.051. This methodological triangulation strengthens the validity of the statistical modeling.

Adding COM to the model proved to be a natural extension. In Finnish upper secondary education, students are accustomed to using electronic textbooks, completing assignments on computers, and utilizing software-based advanced calculators. Both coursework and national exams are conducted digitally \cite{virtanen2024vector}. GenAI represents an additional layer to this established digital learning environment. Familiarity with digital platforms positively influences GenAI's perceived usefulness (PU) and enjoyment (PE).

\subsection{Answers to research questions}

\textbf{(1) To what extent do key constructs of the TAM model influence students' Intention to Use GenAI in general upper secondary mathematics education?} 
The results of this study demonstrated that Perceived Usefulness (PU) was the strongest determinant of students' Intention to Use (ITU) GenAI tools in upper secondary mathematics education. On the other hand, Perceived Ease of Use (PEOU) had a much weaker, non-significant impact on ITU, suggesting that ease of use is less critical when students are focused on achieving tangible academic outcomes. Perceived Enjoyment (PE), while contributing to students' overall perceptions of GenAI, did not have a direct impact on intention to use it. This indicates that Finnish students prioritize the practical benefits of these tools over the enjoyment they experience while using them. 

However, PE and PEOU influenced PU strongly, although the latter was not statistically significant. This is partly aligned with previous study  \cite{Teo2011-zp}, which suggests that PE primarily influences perceived usefulness and ease of use, highlighting that users are more likely to find the technology useful and easy to use when they enjoy the use of it \cite{Teo2011-zp}. Furthermore, the model accounted for 80.4\% of the variance in \textit{ITU}; 60.9\% of the variance in \textit{PU}, and 51.1\% of the variance in \textit{PEOU}, demonstrating strong predictive power for the intention to use GenAI in education. 

\textbf{(2) Are there differences in how these constructs operate in the Finnish educational context compared to findings from similar studies?}
Some differences were observed when comparing the results of this study to a similar study conducted in Hong Kong. In the Finnish context, PU played a much larger role in predicting ITU, suggesting that Finnish students place more emphasis on the practical benefits of GenAI. In contrast, the Hong Kong study showed that PE had a stronger influence on technology adoption, suggesting that intrinsic motivation and enjoyment were more critical for students in that setting. These differences highlight the importance of cultural and educational contexts in shaping how students perceive and adopt new technologies like GenAI. Compared to \cite{strzelecki2023}, using the Unified Theory of Acceptance and Use of Technology (UTAUT), \textit{Performance Expectancy} (PU → ITU) significantly influences students' behavioral intention to use GenAI (0.261, p < 0.001), though the effect was weaker compared to our result. \textit{Effort Expectancy} (PEOU → ITU) also played a modest but significant role (0.085, p = 0.028), while \textit{Hedonic Motivation} (PE → ITU) showed a significant impact on behavioral intention (0.221, p < 0.001). These findings suggest that both perceived usefulness and enjoyment were key drivers for adopting GenAI within Polish higher education, with ease of use being a lesser but still relevant factor.

\textbf{(3) How does the additional variable, Compatibility, affect students' perceptions of using GenAI tools? }
The introduction of Compatibility (COM) enhanced the explanatory power of the model, especially in predicting PU. Students were more likely to find GenAI tools useful when these tools aligned with their prior experiences and the digital tools they were already accustomed to. This may be relevant in the educational context, where students' familiarity with digital learning environments makes GenAI feel more compatible and seamlessly integrated into their existing educational practices. However, the effect of COM on ITU was not as strong, suggesting that while compatibility enhances the perceived usefulness, it does not directly motivate students to adopt the tools.

\section{Discussion}

Cultural differences can affect the adoption of e-learning tools in educational contexts. For example, subjective norms and self-efficacy have a stronger influence in collectivistic cultures, where social pressures and group dynamics shape learners' behavior \cite{zhao2020}. In individualistic cultures, perceived usefulness is the primary determinant for adopting e-learning tools \cite{zhao2020}. Similarly, in the Finnish upper secondary school context, perceived usefulness emerged as a key factor influencing students' intention to use GenAI tools. This highlights the importance of practical utility in the adoption of educational technology. For instance, students view these tools as valuable for improving academic performance, especially in high-stakes settings like exams. In contrast, university students in Hong Kong, who come from diverse cultural backgrounds and are further along in their academic journeys, emphasized enjoyment when interacting with GenAI. 

One possible explanation is that Finnish upper secondary school students aim to compete for positions in further education, whereas Hong Kong university students are already pursuing their goal of graduation. Compared to Hong Kong students, Finnish upper secondary students were under 19 and part of a local, rather homogeneous group, while university students in Hong Kong were typically 19 or older and represented a diverse population, including individuals from across the country and international students. Given that academic success, particularly in preparation for Finnish national exams that serve as a gateway to higher education, it is not surprising that perceived usefulness (PU) strongly influences intention to use (ITU), while perceived ease of use (PEOU) plays a lesser role in this context. However, it is important to note that PE showed a strong connection to PU, suggesting that students may perceive GenAI as both enjoyable and useful, although the direct impact of enjoyment (PE) on ITU is weaker.

Furthermore, one notable difference between Finnish and Hong Kong studies is within path coefficients. In structural equation modeling (SEM), path coefficients represent the strength and direction of the relationships between constructs. According to \cite[e.g.,][]{hair2016primer}, path coefficients are typically standardized values that range between -1 and +1, with higher values indicating stronger relationships between constructs. Path coefficients can be interpreted similarly to standardized beta coefficients in regression analysis: a path coefficient shows how much the endogenous construct changes when the exogenous construct increases by one unit while holding all other factors constant \cite[e.g.,][]{hair2016primer}. In the present study, we observed notably larger path coefficients compared to the Hong Kong study on GenAI adoption in education. For instance, in our analysis, the path coefficients for PU and intention to use (ITU) are notably higher. The larger path coefficients in our study indicate that changes in students' perceptions of usefulness (PU) lead to greater changes in their intention to use (ITU) GenAI tools compared to the Hong Kong study.

Overall, it is notable that the interface of GenAI for mathematics is technically efficient. It can both comprehend mathematical language and generate clear, readable mathematical expressions. Students can input problems as images or through speech, and the operational logic of GenAI is straightforward, allowing for communication as if they were speaking with another person. This communicative feature stands out compared to typical user interfaces of computer or mobile applications. However, students' challenges regarding ease of use (PEOU) seem to be connected to how GenAI can guide them toward the appropriate mathematical methods and tools that are beneficial for them. Moreover, ensuring that GenAI facilitates learning—rather than merely solving the immediate problem—is essential. Therefore, teachers need to support this process by instructing students in pedagogical prompting techniques, which could help optimize the interaction for educational purposes.

\subsection{Limitations and future research}

This study, like all studies, has its limitations. Although the sample size was sufficient for PLS-SEM analysis, it was relatively small, which may limit the generalizability of the findings. However, it is important to note that the sample size reflects this study's aim to explore the immediate effects of using Copilot since its introduction, based on students' responses to the topic. Additionally, the students' trial period with Copilot was brief, suggesting the need for future studies to explore the development of technology acceptance over time in greater depth.

This study intended to reproduce and extend prior work by comparing the results with another study conducted with students in Hong Kong. Despite the intriguing insights, it is important to recognize that there are also differences between these two studies. The sample size in upper secondary school was relatively small, while in the university context, it was considerably large. The students of general upper secondary school studied mathematics, while in the university context, the specific subject was not defined. 

While perceived use has emerged as a critical factor influencing students' intention to use GenAI, an important consideration remains; Is GenAI truly improving learning outcomes, particularly for those students experiencing skill gaps or students with advanced mathematical skills? Does using GenAI lead to a deeper understanding of mathematics or merely an easier way to reach short-term gains while learning mathematics? Especially, for students struggling with foundational math skills, GenAI's ability to bridge these gaps depends on how effectively it facilitates conceptual understanding rather than just providing quick answers. Additionally, further research is needed to explore the impact of GenAI on various student groups, such as those with learning difficulties, to better understand how technology can be effectively optimized to support measurable academic growth. Future studies should also investigate how accepting GenAI and related technologies influences learning outcomes. Furthermore, future research could examine the use of multimodal approaches for GenAI, beyond text-to-text technologies, in mathematics education \citep{Heilala2024-be}. 
 
\section{Conclusion}

In this study, we applied structural equation modeling to investigate students' perceptions of using GenAI in mathematics.  Our study demonstrates that the extended TAM model effectively captured the key factors influencing students' acceptance of generative AI in mathematics education. In the case of upper secondary education in Finland, Perceived Usefulness was the strongest determinant of students' Intention to Use GenAI tools, and Compatibility, in turn, had a significant influence on Perceived Usefulness. Additionally, cultural differences emerged, as Finnish students prioritized the practical benefits of GenAI. At the same time, intrinsic motivation played a larger role in technology adoption among higher education students in Hong Kong \cite{lai2023}. Our study sheds light on those factors and interrelationships that need to be taken into account when using GenAI techniques in mathematics education.

\begin{acks}

Support from the Research Council of Finland under Grant number 353325 is gratefully acknowledged.

\end{acks}

\bibliographystyle{ACM-Reference-Format}
\bibliography{bib} 

\end{document}